\documentclass[12pt, fleqn]{iopart}

\usepackage{iopams}
\usepackage{setstack}

\usepackage{graphicx}
\usepackage{theorem}
\usepackage{hyperref}

\theoremstyle{plain}
\newtheorem{Thm}{Theorem}

\newtheorem{Cor}[Thm]{Corollary}
\newtheorem{Prop}[Thm]{Proposition}

\theorembodyfont{\rmfamily}
\newtheorem{Proof}{Proof}

\begin{document}

\title{The solution to the initial value problem for the ultradiscrete Somos-4 and 5 equations.}
\author{Yoichi Nakata}
\address{$^1$ Interdisciplinary Center for Mathematical Sciences, Graduate School of Mathematical Sciences, the University of Tokyo, 3-8-1 Komaba, Meguro-ku, 153-8914 Tokyo, Japan}
\address{$^2$ Institute for Biology and Mathematics of Dynamical Cell Processes (iBMath), the University of Tokyo, 3-8-1 Komaba, Meguro-ku, 153-8914 Tokyo, Japan}
\ead{ynakata@ms.u-tokyo.ac.jp}
\begin{abstract}
We propose a method to solve the initial value problem for the ultradiscrete Somos-4 and Somos-5 equations by expressing terms in the equations as convex polygons and regarding max-plus algebras as those on polygons.
\end{abstract}
\vspace{2pc}

\section{Introduction}

It is still a difficult problem to define the integrability of discrete equations in a way that does not rely on the properties differently from that of differential ones. Several criteria have been proposed for solving this problem by observing the behavior of the solutions to discrete equations which are considered as integrable ones. For example, in the singularity confinement test \cite{GrammaticosRamaniPapageorgiou}, the property that the singularities due to an initial value are resolved after  several time steps and that the information on the initial value is finally restored, is considered to be a discrete analogue of the Painlev\'e property, which is an indication of integrability. The algebraic entropy \cite{BellonViallet} focuses on the growth of the degree of the solution as a rational expression of the initial values. It is considered that the system is integrable if the degree grows in at most polynomial order and is non-integrable if the order is exponential. These criteria are also related to the structure of discrete equations such as co-primeness and irreducibility \cite{KankiMadaMaseTokihiro}.

Over the past decade, it was discovered that cluster algebras, introduced by Fomin and Zelevinsky \cite{FominZelevinsky1}, are strongly related with discrete integrable equations \cite{InoueIyamaKunibaNakanishiSuzuki, Okubo1}. The time evolution of many integrable discrete equation can be expressed as the mutation of cluster variables, where these cluster variables are expressed not as rational expressions but in the form of Laurent polynomials of the initial values by properly performing fractional reduction in the recursive application of the equation (which includes  divisions \cite{FominZelevinsky2}). Furthermore, recent studies discovered that such polynomials are irreducible and co-prime for known integrable discrete equations and these properties correspond to the criteria described above \cite{KankiMadaMaseTokihiro}.

Ultradiscrete systems are difference equations in which only $\max$ and $\pm$ operators appear. These equations are obtained from minus-free canonical difference equations by a limiting procedure called ``ultradiscretization" \cite{TokihiroTakahashiMatsukidairaSatsuma}, which is defined as follows:
\begin{enumerate}
\item Transform the dependent variables and parameters by exponential functions, upon introduction of a positive parameter $\varepsilon$, for example $a=e^{A/\varepsilon}$ where $a$ is the dependent variable or the parameter in the discrete system.
\item Take the logarithm of each side of the equation and take the limit $\varepsilon \to +0$. Then, by means of the identity
\begin{equation}
	\lim_{\varepsilon \to +0} \varepsilon \log ( e^{A/\varepsilon} + e^{B/\varepsilon} ) = \max (A, B)
\end{equation}
and the exponential law, the operators $+$ and $\times$ in canonical difference equations are replaced with $\max$ and $+$ respectively.
\end{enumerate}
The remarkable point of this procedure is that it preserves the good properties of integrable systems, although the dependent variables only take discrete values. The most famous example is the Box and Ball system (BBS) \cite{TakahashiSatsuma}, which is a cellular automaton consisting of an infinite sequence of boxes and a finite amount of balls. The BBS has solitons and an infinite amount of conserved quantities and is obtained by the ultradiscretization of the KdV equation.

For the ultradiscrete equations, we can obtain solutions by ultradiscretizing those of discrete equations. However, it still remains the problem how to interpret good properties of the equation, for example, the Laurent phenomenon in the ultradiscrete systems. By ultradiscretizing Laurent polynomials naively, one expects that the form of solutions should be expressed as $\max_{i=1, \ldots, N} (F_i(\mathbf{A}))$, where $F_i$ is a linear function of $\mathbf{A} \in \mathbb{R}^n$. However, a mechanism corresponding to the reduction of the fraction is required in the operation to keep such a form even if the evolution equation contains minus terms. We believe that such a mechanism can be explained by using combinatorics and we finally conclude that it can be interpreted as the inverse of the Minkowski sum between convex polygons. Applying this idea to several known integrable ordinary difference equations, we obtain the exact solution to their initial value problems.

In this paper, we first explain this key idea by a simple ultradiscrete equation in Section 2. By virtue of this idea, we introduce the solution of the initial value problem to the ultradiscrete Somos-4 equation and discuss properties of its solutions and the relation with an ultradiscrete QRT map in Section 3. We also introduce the solution to ultradiscrete Somos-5 equation in Section 4.

\section{Key idea}
Let us consider the following equation, which arises as the mutation of cluster variables in an $A^{(1)}_1$-type cluster algebra
\begin{equation}
	f_n f_{n-2} = f_{n-1}^2 + 1 \quad (n \ge 2). \label{da11}
\end{equation}

Here, as an evolution equation (\ref{da11}) contains a division. However, $f_n$ is always a Laurent polynomial of $f_0$ and $f_1$ with positive coefficients \cite{CalderoZelevinsky}. Therefore, if the initial values $f_0$ and $f_1$ are positive, all $f_n$ take positive values and ultradiscretizable in the sense of \cite{TokihiroTakahashiMatsukidairaSatsuma}. Applying the ultradiscretization procedure to (\ref{da11}), we obtain:
\begin{equation}
	F_n + F_{n-2} = 2\max(F_{n-1}, 0) \quad (n \ge 2). \label{uda11}
\end{equation}
Due to the Laurent phenomenon for the discrete system, the solution to the ultradiscrete system (\ref{uda11}) should be expressible as:
\begin{equation}
	F_n = \max_{(\alpha, \beta) \in V_n} (\alpha A + \beta B), \label{FnLP}
\end{equation}
where $V_n \subset \mathbb{Z}^2$ is a finite set, $A = F_0$ and $B=F_1$. On the other hand, the equation (\ref{uda11}) can behave a evolution equation, that is, we can obtain $F_n$ uniquely by recurrence. 

For example, the solution $F_n$ for the first several $n$ is obtained as
\begin{eqnarray}
  	F_2 = 2\max(B, 0) - A = \max(-A+2B, -A)\\
	F_3 = 2\max(2B, 0, A) - 2A -B = \max(-2A+B, -2A-B, -A-B) \\
	F_4 = 2\max(4B, 0, 2A, 2A+B) - 2\max(B, 0) - 3A-2B. \label{F4}
\end{eqnarray}
Here, by virtue of the rules of the $\max$-plus algebra, one has
\begin{eqnarray}
       \max(4B, 0, 2A, 2A+B) &= \max(4\max(B, 0), 2A+\max(B, 0)) \nonumber\\
       					  &= \max(3\max(B, 0), 2A) + \max(B, 0) \nonumber\\
					  &= \max(3B, 0, 2A) + \max(B, 0).
\end{eqnarray}
Then, $-\max$ in (\ref{F4}) is cancelled and it is finally simplified into
\begin{equation}
	F_4 = 2\max(3B, 0, 2A) - 3A - 2B = \max(-3A+4B, -3A-2B, A-2B).
\end{equation}
Continuing the calculation, we obtain
\begin{equation}
	 F_5 = 2\max(6B, 0, 4A, 3A+2B) - 2\max(2B, 0, A) - 4A - 3B. \label{F5}
\end{equation}
However, in this case there is no immediately apparent way to put formula (\ref{F5}) in the form (\ref{FnLP}), which should nonetheless be feasible because of the uniqueness of the solution to the evolution equation (\ref{uda11}).
Analyzing the right-hand side of (\ref{F5}) case by case, we can simplify $F_5$ into
\begin{equation}
	F_5 = 2\max(4B, 0, 3A) - 4A - 3B.
\end{equation}
Therefore, the following identity should hold in general:
\begin{equation}
     \max(4B, 0, 3A) + \max(2B, 0, A) = \max(6B, 0, 4A, 3A+2B).
\end{equation}
Our goal is to explain this identity by means of a general procedure. By naively expanding the left hand side, we obtain
\begin{equation}
      \max(4B, 0, 3A) + \max(2B, 0, A) = \max(6B, 4B, A+4B, 2B, 0, A, 3A+2B, 3A, 4A). \label{maxidentity}
\end{equation}
Therefore, to prove the identity one has to show that $4B$, $A+4B$, $2B$, $A$, $3A$ are less than $\max(6B, 0, 4A, 3A+2B)$. Here, $A+4B$ can be expressed as
\begin{equation}
	A+4B = \frac{1}{4} \times 4A + \frac{2}{3} \times 6B + \frac{1}{12} \times 0
\end{equation}
and the summation of coefficients of $4A$, $6B$ and $0$ is $1$, that is, $A+4B$ is written in a convex combination of $4A$, $6B$ and $0$. It is trivial to see that other terms are also written as convex combinations. We can evaluate the magnitude relationship for such convex combined terms by the following proposition.

\begin{Prop}
For the finite set of points $\{ (\alpha_i, \beta_i) \}_{i=1}^M \subset \mathbb{R}^2$, if there exists $j \in \{ 1, $\ldots$, M \}$ satisifying 
\begin{equation}
    (\alpha_j, \beta_j) = \sum_{\scriptsize \begin{array}{c} $i=$1 \\ $i$\neq$j$ \end{array}}^M \lambda_i (\alpha_i, \beta_i) \label{wm}
\end{equation}
for some $\sum_{i=1, i \neq j}^M \lambda_i = 1$, $\lambda_i \ge 0$, one has
\begin{equation}
    \max_{i=1, \ldots, M}(\alpha_i A + \beta_i B) = \max_{\scriptsize \begin{array}{c} $i=$1, \ldots, M \\ $i$\neq$j$ \end{array}}(\alpha_i A + \beta_i B).
\end{equation}
\end{Prop}

\begin{Proof}
By virtue of equation (\ref{wm}), one has
\begin{equation}
	\alpha_j A + \beta_j B = \sum_{\scriptsize \begin{array}{c} $\it{i=}$1 \\ $\it{i}$\neq$\it{j}$ \end{array}}^M \lambda_i (\alpha_i A + \beta_i B),
\end{equation}
which means $(\alpha_j, \beta_j)$ is expressed as the weighted average of other $(\alpha_i, \beta_i)$, i.e., it is less than the maximum of others and more than the minimum. \hfill $\square$
\end{Proof}

With this proposition, it is easily confirmed that (\ref{maxidentity}) holds. Now, let us proceed further with this proposition.

\begin{Cor}
Let $V = \{(\alpha_i, \beta_i)\}_{i=1}^N$ and let $V_e \subset V$ be the set of extreme points of $V$ (the vertices of the convex hull of $V$), then
\begin{equation}
  	\max_{(\alpha, \beta) \in V}(\alpha A + \beta B) = \max_{(\alpha, \beta) \in V_e}(\alpha A + \beta B).
\end{equation}
\end{Cor}

\begin{Prop}
For all $(\alpha', \beta') \in V_{e}$, there exists $(A, B) \in \mathbb{R}^2$ such that $\alpha' A + \beta' B > \max_{(\alpha,\beta) \in V_{e} \backslash \{ (\alpha', \beta') \}} (\alpha A + \beta B)$, that is, $\max_{(\alpha,\beta) \in V_{e} \backslash \{ (\alpha', \beta') \}} (\alpha A + \beta B) \neq \max_{(\alpha,\beta) \in V_{e}} (\alpha A + \beta B)$.
\end{Prop}

\begin{Proof}
By assumption, the points of $V_e$ are the vertices of a convex polygon. Let $N = \#V_e$ and let $e_i$ ($i=1, \ldots, N$) be an element of $V_e$ where the vertices are ordered by counter-clockwise and $n_i := e_{i+1}-e_i$ ($i=1, \ldots, N-1$) and $n_N := e_1 - e_N$ are the edge vectors of the polygon. It is sufficient to prove that only $\alpha_1 A + \beta_1 B$ attains the maximum of $\max_{i=1, \ldots, N} (\alpha_i A + \beta_i B)$ in some region. Let us consider the open cone $\{ x = (A, B) \in \mathbb{R}^2 \mid \langle n_1, x \rangle > 0, \langle n_N, x \rangle < 0 \}$. In this cone, one has $\alpha_2 A + \beta_2 B < \alpha_1 A + \beta_1 B$ and $\alpha_N A + \beta_N B < \alpha_1 A + \beta_1 B$. For other $\alpha_i A + \beta_i B$, the specific magnitude relationship will change depending on where $(A, B)$ is in the cone, but in any case it is finally proved that it is less than $\max(\alpha_1 A + \beta_1 B, \alpha_N A + \beta_N B)$. \hfill $\square$
\end{Proof}

\begin{Cor}
Let $\mathcal{F} = \{ f: \mathbb{R}^2 \to \mathbb{R} \mid f(A, B) = \max_{(\alpha, \beta) \in V} (\alpha A + \beta B), \mbox{$V \subset \mathbb{R}^2$ is finite set.} \}$. Then, there exists one-to-one correspondence between convex polygons on $\mathbb{R}^2$ and elements of $\mathcal{F}$.
\end{Cor}

By this corollary, we can regard formulae for $\max$ as convex polygons. Next we want to interpret the algebra for $\max$ formulae as polygon operations. By the relations
\begin{eqnarray}
	\max(\max_i(\alpha_i A + \beta_i B), \max_j(\gamma_j A + \delta_j B)) = \max_{i, j}(\alpha_i A + \beta_i B, \gamma_j A + \delta_j B) \\
	\max_i (\alpha_i A + \beta_i B) + \max_j (\gamma_j A + \delta_j B) = \max_{i, j}((\alpha_i+\gamma_j)A + (\beta_i+\delta_j) B),
\end{eqnarray}
we obtain that $\max$ operation gives the convex hull of the union of two polygons and $+$ operation gives the Minkowski sum of two polygons, where the Minkowski sum of two subsets is defined as $U+V := \{ u+v \mid u \in U, v \in V \}$.

From these dicussions, it is found that the expressions of $\max$ correspond to convex polygons and the $\max$-plus algebra for these expressions can be replaced with calculations on convex polygons. In general, however, it is very difficult to determine the extreme points of the Minkowski sum. Fortunately, by virtue of the results of computational geometry, there is a simple method to calculate Minkowski sums for planar convex polygons, by focusing on their edges \cite{EmirisTsigaridas}.

\begin{Prop}(\cite{EmirisTsigaridas})
Let $P$, $Q$ be convex polygons in $\mathbb{R}^2$ and let $E(X)$ be the set of edge vectors of polygon $X$. Then, the edges of their Minkowski sum $E(P+Q)$ are obtained by the following algorithm:

\begin{itemize}
\item Let $E(P) = \{ e_i \}_{i=1}^n$, $E(Q) = \{ \tilde{e}_j \}_{j=1}^m$, where indices are sorted by the argument.
\item Start from $i=1$ and $j=1$ and apply the following until $i>n$ or $j>m$:
\item Compare two arguments of $e_i$ and $\tilde{e}_j$.
\begin{itemize}
\item If $\arg e_i > \arg \tilde{e}_j$, append $e_i$ to $E(P+Q)$ and let $i \mapsto i+1$.
\item If $\arg e_i < \arg \tilde{e}_j$, append $\tilde{e}_j$ to $E(P+Q)$ and let $j \mapsto j+1$.
\item If $\arg e_i = \arg \tilde{e}_j$, append $e_i + \tilde{e}_j$ to $E(P+Q)$ and let $i \mapsto i+1$ and $j \mapsto j+1$.
\end{itemize}
\item If $i>n$, append $\tilde{e}_j, \ldots, \tilde{e}_m$ to $E(P+Q)$.
\item If $j>m$, append $e_i, \ldots, e_n$ to $E(P+Q)$.
\end{itemize}
\end{Prop}
We note that $\max(0, A)$ does not seems to be a polygon but a line segment. In this case, we consider this as a dihedral and its edge vectors are $\{(1, 0), (-1, 0)\}$. We also note that the sum of all edge vectors is $0$.

Here, we demonstrate this algorithm by an example. Let us consider two polygons $P=\{ (0, 4), (0, 0), (3, 0) \}$ and $Q = \{ (0, 2), (0, 0), (1, 0) \}$ (we express polygons by their extreme points) . The edge vectors of each polygon are expressed as $E(P) = \{ (0, -4), (3, 0), (-3, 4) \}$ and $E(Q)=\{ (0, -2), (1, 0), (-1, 2) \}$. Then, the edge vectors of their Minkowski sum are $E(P+Q)=\{ (0, -6), (4, 0), (-1, 2), (-3, 4) \}$. By transforming this to extreme points, one has $P+Q = \{ (0, 6), (0, 0), (4, 0), (3, 2) \}$, which is another proof of identity (\ref{maxidentity}). We can confirm the result visually in Figure 1.

\begin{figure}
\begin{center}
\includegraphics{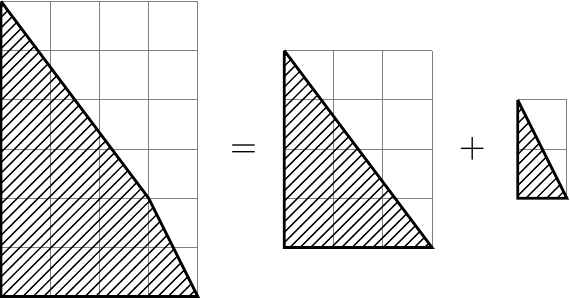}
\end{center}
\caption{Polygon interpretation of equation (\ref{maxidentity}). The $+$ operator in the $\max$-plus algebra corresponds to the Minkowski sum of polygons.}
\end{figure}

The remarkable point is that we can obtain the inverse of the Minkowski sum by executing this algorithm, which yields that the necessary and sufficient condition for calculating the inverse of the Minkowski sum is that there are no edge vectors which are contained by the subtrahend polygon and not contained by the minuend one.

By virtue of these discussions, we can regard the max-plus algebra as polygon calculus and apply this result to ultradiscrete equations which correspond to discrete ones that have the Laurent property.

\bigskip

For example, let us go back to obtain the solution $F_n$ of (\ref{uda11}). Because of the discussions above, $F_n$ can be solved and written as
\begin{equation}
  	F_n = 2\max\Big((n-1)B, 0, (n-2)A\Big) - (n-1)A - (n-2)B \quad (n \ge 2) \label{uda11sol}
\end{equation}
because of the identity
\begin{equation}
    \!\!\!\!\!\!\!\!\!\!\!\! \max\Big(0, nA, (n+1)\Big) + \max\Big(0, (n-2)A, (n-1)B\Big) = \max\Big(0, 2(n-1)A, 2nB, nA + (n-1)B\Big)
\end{equation}
for $n \ge 2$.

Equation (\ref{da11}) is equivalent to the linear equation:
\begin{equation}
	f_n + f_{n-2} = \frac{f_0^2 + f_1^2 + 1}{f_0 f_1} f_{n-1}. \label{da11lin}
\end{equation}
By ultradiscretizing this relation, we find
\begin{equation}
	\max(F_n, F_{n-2}) = F_{n-1} + 2\max(A, B, 0) - A - B. \label{uda11lin}
\end{equation}
Substituting (\ref{uda11sol}), we have another identity
\begin{eqnarray}
	\max\Big((n-1)B, 0, (n-2)A, (n-2)B+A, A+B, (n-3)A+B \Big) \nonumber\\
		= \max\Big((n-2)B, 0, (n-3)A\Big) + 2\max(B, 0, A).
\end{eqnarray}
for $n \ge 2$ which can be proven using polygon calculus. It is an interesting point that we cannot obtain $F_n$ recursively from (\ref{uda11lin}) because it does not form an evolution equation although the corresponding discrete equation (\ref{da11lin}) is linear (which is generally considered to be easier to solve than a non-linear one).

\bigskip

The polygon corresponding to the $\max$ formula is nothing but the Newton Polygon of the polynomial before the ultradiscretization. It is known that the Newton Polygon behaves as a lattice for the union and Minkowski sum operation. One can obtain that the necessary condition to factorize a polynomial  is that its Newton polygon is decomposable. However, this is not sufficient. For example, $a^2+3a b + b^2 \neq (a+b)^2$ but their ultradiscretizations are equal. Furthermore, the polygon decomposition is not unique. For example, $\max(2E, 2D, 3D, 3D+E, 3D+E, D+2E)$ is decomposed into $\max(E, D, D+E) + \max(E, D, 2D)$ or $\max(D, 0) + \max(2E, 2D, 2D+E)$ and both of these cannot be decomposed further. The reason for such phenomena is that the Newton polygon ignores the terms except those corresponding to the extreme points.

Finally, let us note that the polygon we dealt with in this section is considered to be dual to a tropical curve, and that operations between polygons can therefore be also interpreted as operations on tropical curves.

\section{Ultradiscrete Somos-4}
The Somos sequences are the difference equations expressed as
\begin{equation}
	f_n f_{n-q} = \sum_{i=1}^{\lfloor \frac{q}{2} \rfloor} f_{n-i} f_{n-q+i},
\end{equation}
where $q$ is an integer more than 4. This equation is also called Somos-$q$ equation for some specific value of $q$.

In these sequences, the case where $q$ satisfies $4 \le q \le 7$ is related to integrable systems.
For these value of $q$, it has been proven that $f_n$ is a Laurent polynomial of $f_0$, $\ldots$, $f_{q-1}$ with positive coefficients \cite{FominZelevinsky2}. Furthermore, these sequences are derived as reductions of some integrable partial difference equations. For example, the Somos-4 and 5 equations are derived from the discrete KP equation and the Somos-6 and 7 equations are from the discrete BKP equation \cite{Hone}. We also note that Somos-6 and 7 are not obtained from cluster algebras rather from Laurent Phenomenon algebras \cite{Okubo2}, which are analogues of the cluster algebras \cite{LamPylyavskyy}.

By applying the ultradiscretization procedure to the Somos-4 equation, we obtain
\begin{equation}
	F_n + F_{n-4} = \max(F_{n-1} + F_{n-3}, 2F_{n-2}) \quad (n \ge 4). \label{udsomos4}
\end{equation}
We call this the ultradiscrete Somos-4 equation. This equation is a fourth order difference equation and solutions are expressed in terms of $F_0$, $F_1$, $F_2$, $F_3$. However, since this equation is invariant under the gauge transformation $F_n \mapsto F_n + a + b n$ ($a, b \in \mathbb{R}$), we can set $F_0=F_1=0$ without loss of generality by taking the proper gauge. Therefore, the solutions are expressed by a planar polygon. Applying the ideas of the previous section to this equation, we can solve its initial value problem.

\begin{Thm}
The solution of the initial value problem for (\ref{udsomos4}) is expressed as:
\begin{equation}
	F_n = - \nu_{n+2} C - \nu_{n+1} D + \tilde{F}_n, \label{udsomos4sol}
\end{equation}
where $C=F_2$, $D=F_3$ and $\nu_n$ is the solution to the same equation (\ref{udsomos4}) for the initial values $\nu_0 = 1$, $\nu_1 = \nu_2 = \nu_3 = 0$, represented as follows:
\begin{eqnarray}
	\nu_{8k} = 4k^2 - 4k + 1 \label{udsomosspsol0} \\
	\nu_{8k+1} = 4k^2 - 3k \\
	\nu_{8k+2} = 4k^2 - 2k \\
	\nu_{8k+3} = 4k^2 - k \\
	\nu_{8k+4} = 4k^2 -1\\
	\nu_{8k+5} = 4k^2 + k \\
	\nu_{8k+6} = 4k^2 + 2k \\
	\nu_{8k+7} = 4k^2 + 3k. \label{udsomosspsol7}
\end{eqnarray}
$\tilde{F}_n$ is
\begin{eqnarray}
	\tilde{F}_{8k} = (4k^2-k) Q + k P \\
	\tilde{F}_{8k+1} = 4k^2 Q + k P \\
	\tilde{F}_{8k+2} = (4k^2+k) Q + k P \\
	\tilde{F}_{8k+3} = (4k^2+2k) Q + k P \\
	\tilde{F}_{8k+4} = (4k^2+3k) Q + k P + \max(D, 2C) \\
	\tilde{F}_{8k+5} = (4k^2+4k) Q + k P + \max(D, C+D, 3C) \\
	\tilde{F}_{8k+6} = (4k^2+5k) Q + k P + \max(3D, 2D, 4C, 3C+D) \\
	\tilde{F}_{8k+7} = (4k^2+6k) Q + k P + \max(4D, 3D, 6C, 7C, C+4D),
\end{eqnarray}
and $Q = \max(2D, D, 2C, 3C)$ and $P = \max(4D, 3D, 6C)$.
\end{Thm}

Before starting the proof, we calculate the first several expressions by the recurrence and obtain
\begin{eqnarray}
	F_0 = F_1 = 0, F_2 = C, F_3 = D \\
	F_4 = \max(D, 2C) \\
	F_5 = \max(D, C+D, 3C) \\
	F_6 = -C + \max(3D, 2D, 4C, 3C+D) \\
	F_7 = -D + \max(4D, 3D, 6C, 7C, C+4D),
\end{eqnarray}
which are consistent with the above result for $k=0$.

\begin{Proof}
We first prove the statement concerning $\nu_n$. Substituting $\nu_n$ in both sides of (\ref{udsomos4}), the terms depending on $k$ are factored out from the $\max$ in the right hand side, such that the terms are the same on both sides. For example, $\nu_{8k+4}+\nu_{8k} = 8k^2 - 4k$ and $\max(\nu_{8k+3} + \nu_{8k+1}, 2\nu_{8k+2}) = 4k^2-4k$. Therefore, we should consider only the cases from $n=4$ to $11$ and prove these by simple calculations. 

Next we consider $\tilde{F}_n$. Since $\nu_n$ satisfies (\ref{udsomos4}), we transform (\ref{udsomos4}) to an equation for $\tilde{F}_n$:
\begin{equation}
	\tilde{F}_n + \tilde{F}_{n-2} = \max(\tilde{F}_{n-1} + \tilde{F}_{n-3} + (d_n)_{-} C + (d_{n-1})_{-}D, 2\tilde{F}_{n-2} + (d_n)_{+}C + (d_{n-1})_{+}D), \label{udsomos4mod}
\end{equation}
where $d_n := \nu_n + \nu_{n+2} - 2\nu_{n+1}$, $(a)_{+} := \max(a, 0)$ and $(a)_{-} := \max(-a, 0)$. Here, due to (\ref{udsomosspsol0})--(\ref{udsomosspsol7}), one obtains that $d_n$ has period $8$.

Therefore, by virtue of the same discussion as for $\nu_n$ solving (\ref{udsomos4}), we have to consider only the cases from $n=4$ to $11$ and obtain that $\tilde{F}_n$ solves (\ref{udsomos4mod}) by the using polygon calculus in the previous section.
\hfill $\square$
\end{Proof}

Now, we focus on the properties of the solution that we obtained. We first point out that the solution (\ref{udsomos4sol}) is decomposable (actually already decomposed) and contains the same polygon in decomposed ones in contrast with the irreducibility and co-primeness of the solution which was proven for the (discrete) Somos-4 equations. The reason is the same as for the polygon expression described in the previous section. We also note that the growth of the coefficients of $C$, $D$ in the solution for $n$ is of square order, which follows the preceding studies \cite{Mase}.

We next discuss the relation to the QRT systems. By introducing the dependent variable $g_n = f_n f_{n+2}/f_{n+1}^2$, the Somos-4 is written as
\begin{equation}
	g_n g_{n-2} = \frac{g_{n-1}+1}{g_{n-1}^2}, \label{dQRT}
\end{equation}
which is one of the QRT maps \cite{Hone}. The corresponding ultradiscrete dependent variable is
\begin{equation}
	G_n = F_n + F_{n+2} - 2F_{n+1} \label{somosQRT}
\end{equation}
and the ultradiscrete Somos-4 (\ref{udsomos4}) is transformed into
\begin{equation}
	G_n + G_{n-2} = \max(G_{n-1}, 0) - 2G_{n-1}, \label{udQRT}
\end{equation}
which is one of the ultradiscrete QRT maps.

By substituting (\ref{udsomos4sol}) into relation (\ref{somosQRT}) we obtain the following corollary.

\begin{Cor}
The solution to the equation (\ref{udQRT}) for the initial values $G_0 = C$ and $G_1 = -2C+D$ is expressed as
\begin{eqnarray}
	G_{8k} = C \\
	G_{8k+1} = -2C+D \\
	G_{8k+2} = C-2D + \max(D, 2C) \\
	G_{8k+3} = D + \max(2D, C+D, 3C) - 2\max(D, 2C) \\
	G_{8k+4} = -C + \max(3D, 2D, 4C, 3C+D) + \max(D, 2C) \nonumber\\
	\quad -2\max(2D, C+D, 3C) \\
	G_{8k+5} = C-D+\max(4D, 3D, 6C) + \max(0,C) \nonumber\\
	\quad + \max(2D, C+D, 3C) - 2\max(3D, 2D, 4C, 3C+D) \\
	G_{8k+6} = -C+D + \max(3D, 2D, 4C, 3C+D) + \max(2D, D, 2C, 3C) \nonumber\\ 
	\quad - \max(4D, 3D, 6C) - 2\max(0, C) \\
	G_{8k+7} = -D + \max(0, C).
\end{eqnarray}
Therefore, the period of the solution is $8$ for arbitrary initial values.
\end{Cor}
This corollary can be also proved by directly calculating $G_n$ from equation (\ref{udQRT}) recurrently.

Nobe solved the ultradiscrete QRT maps including (\ref{udQRT}) by regarding the systems as additions on Tropical Elliptic Curves and obtained the same result \cite{Nobe}. In \cite{Nobe} it is pointed out that the solution to the discrete equation (\ref{dQRT}) has no periodicity, although that to the ultradiscrete equation (\ref{udQRT}) is periodic. The reason why the discrete equation has no periodicity is explained by the irreducibility and co-primeness of the solution \cite{KankiMadaMaseTokihiro} and by due to the discussion in this section, we must conclude that the ultradiscrete solution has periodicity because such properties are broken by the ultradiscretization. We finally stress that these preceding studies \cite{Nobe, OrmerodYamada} also employ the polygon geometry. However, in their approach, the solution is expressed as a point on polygon facets and our approach considers the solution as a polygon itself, which is a major difference.

\section{Ultradiscrete Somos-5}

By ultradiscretizing the Somos-5 equation, one obtains the ultradiscrete Somos-5
\begin{equation}
	F_n + F_{n-5} = \max(F_{n-1} + F_{n-4}, F_{n-2} + F_{n-3}). \quad (n \ge 5) \label{udsomos5}
\end{equation}
This equation is a fifth order difference equation. However, by employing the same approach to solving the ultradiscrete Somos-4, this equation is invariant under the gauge $F_n \mapsto F_n + a + b n + c (-1)^n$ ($a, b, c \in \mathbb{R}$) and we can set $F_0=F_1=F_2=0$ without loss of generality. Therefore, the solution is also expressed as a planar polygon. Since the approach of the proof is the same as that for the ultradiscrete Somos-4 equation, we omit the details and show only results.
 
\begin{Thm}
The solution is written as
\begin{equation}
	F_n = - \nu_{n+3} D - \nu_{n+2} E + \tilde{F}_n,
\end{equation}
where $D=F_2$, $E=F_3$, $\nu_n$ is the solution to the same equation (\ref{udsomos5}) for the initial value $\nu_0 = 1$, $\nu_1 = \nu_2 = \nu_3 = 0 = \nu_4 = 0$ and represented as follows:
\begin{eqnarray}
	\nu_{7k} &= 1 + \frac{1}{8} (-\phi_k - 6k + \psi_k)  \\
	\nu_{7k+1} &= \frac{1}{8} (\phi_k - 2k + \psi_k) \\
	\nu_{7k+2} &= \frac{1}{8} (-\phi_k + 2k + \psi_k) \\
	\nu_{7k+3} &= \frac{1}{8} (\phi_k + 6k + \psi_k) \\
	\nu_{7k+4} &= \frac{1}{8} (-\phi_k + 10k + \psi_k) \\
	\nu_{7k+5} &= -1 + \frac{1}{8} (\phi_k + 14k + \psi_k) \\
	\nu_{7k+6} &= \frac{1}{8} (-\phi_k + 18k + \psi_k).
\end{eqnarray}
$\tilde{F}_n$ is
\begin{eqnarray}
	\tilde{F}_{7k} = \frac{1}{8} (-\phi_k-2k+\psi_k) Q + \frac{-\phi_k+2k}{4} R + k P \\
	\tilde{F}_{7k+1} = \frac{1}{8} (\phi_k+2k+\psi_k) Q + \frac{\phi_k+2k}{4} R + k P \\
	\tilde{F}_{7k+2} = \frac{1}{8} (-\phi_k+6k+\psi_k) Q + \frac{-\phi_k+2k}{4} R + k P \\
	\tilde{F}_{7k+3} = \frac{1}{8} (\phi_k+10k+\psi_k) Q + \frac{\phi_k+2k}{4} R + k P \\	
	\tilde{F}_{7k+4} = \frac{1}{8} (-\phi_k+14k+\psi_k) Q + \frac{-\phi_k+2k}{4} R + k P \\
	\tilde{F}_{7k+5} = \frac{1}{8} (\phi_k+18k+\psi_k) Q + \frac{\phi_k+2k}{4} R + k P + \max(E, D) \\	
	\tilde{F}_{7k+6} = \frac{1}{8} (-\phi_k+22k+\psi_k) Q + \frac{-\phi_k+2k}{4} R + k P + 	\max(E, D, D+E),
\end{eqnarray}
$\phi_k = 1-(-1)^k$, $\psi_k = 14k(k-1)$, $Q = \max(E, 0) + \max(E, D, 2D)$, $R= \max(E, 0)$ and $P = \max(2E, 2D, 2D+E)$.
\end{Thm}

By introducing a new dependent variable $g_n = f_n f_{n+3}/f_{n+1} f_{n+2}$, the Somos-5 equation can be written as
\begin{equation}
	g_n g_{n-2} = \frac{g_{n-1}+1}{g_{n-1}},
\end{equation}
which is also a QRT map. The corresponding transformation of the dependent variable in the ultradiscrete system is
\begin{equation}
G_n = F_n + F_{n+3} - F_{n+1} - F_{n+2}
\end{equation}
and we obtain its ultradiscretization:
\begin{equation}
	G_n + G_{n-2} = \max(G_{n-1}, 0) - G_{n-1}. \label{udQRT2}
\end{equation}

\begin{Cor}
The solution to the equation (\ref{udQRT}) for the initial values $G_0 = D$ and $G_1 = -D+E$ is expressed as
\begin{eqnarray}
	\!\!\!\! G_{7k} = D \\
	\!\!\!\! G_{7k+1} = - D + E \\
	\!\!\!\! G_{7k+2} = - D - E +\max(E, D) \\
	\!\!\!\! G_{7k+3} = D - E + \max(D, E, D+E) - \max(E, D) \\
	\!\!\!\! G_{7k+4} = E + \max(2E, 2D, 2D+E) - \max(E, D) - \max(E, D, D+E) \\
	\!\!\!\! G_{7k+5} = -D + \max(E, D) - \max(E, D, D+E) + \max(E, 0) \\
	\!\!\!\! G_{7k+6} = -E + \max(E, D, D+E) + \max(E, D, 2D) - \max(2E, 2D, D+E).
\end{eqnarray}
Therefore, the period of the solution is $7$ for arbitrary initial values \cite{Nobe}.
\end{Cor}

\section{Concluding Remarks}

In this paper, we proposed a purely ultradiscrete calculus-based method to solve the initial value problem for the ultradiscrete Somos-4 and 5 equations by regarding $\max$ formulae as convex polygons. The solution can be written as a single $\max$ expression even if the evolution equations contain minus terms, which is an analogue of the Laurent property in ultradiscrete systems.

The idea discussed in Section 2 faithfully replaces the max-plus algebra with polygon operations. This means that problems arising in the max-plus algebra, are also present in polygon operations. For example, by setting $P = \max(B, A, B+2A, A+2B)$ and $Q = \max(0, 2A, 2A+2B, 2B)$, $\max(P, Q)$ in fact no longer depends on $P$, which corresponds to the fact that a polygon included in other polygons, no longer influences their geometrical properties. 

Our approach can be applied to equations with the Laurent property, even if they are higher order ones or have higher degree non-linearities. For example, the equation
\begin{equation}
	f_n f_{n-2} = (f_{n-1})^m + 1 
\end{equation}
for $m > 2$ has the Laurent property. The corresponding ultradiscrete equation is 
\begin{equation}
	F_n + F_{n-2} = m \max(F_{n-1}, 0)
\end{equation}
and its solution is expressed as $F_n = \max(b_{n-1} F_1, b_{n-2} F_0, 0) - \alpha_n F_0 - \alpha_{n-1} F_1$ ($n \ge 2$), where $b_n$ is the solution of $b_n = m b_{n-1} - b_{n-2}$, $b_0 = 0$, $b_1 = m$ and $\alpha_n$ is the solution of the same difference equation $\alpha_n = m \alpha_{n-1} - \alpha_{n-2}$ with the different initial value $\alpha_0 = -1$, $\alpha_1 = 0$. Though the growth of $b_n$ and $\alpha_n$ is of exponential order, the ultradiscretized solution is also expressible by means of the polygons. This solution also holds even when $m$ is not integer. This result may indicate a suggestion on what is the Laurent property about difference equations with the non-integer degree non-linearity.

We finally note that the method used in the proofs of Theorems 6 and 8 to obtain the general solution $F_n$ after finding a special solution $\nu_n$, is very similar to the quadrature method for the general solution of the Riccati equation.

\section*{Acknowledgment}
The author would like to thank Professors T. Tokihiro and R. Willox and Dr. T. Mase for helpful comments. This work was supported by Platform for Dynamic Approaches to Living System (the Platform Project for Supporting in Drug Discovery and Life Science Research) from the Ministry of Education, Culture, Sports, Science (MEXT) and Technology, Japan, and Japan Agency for Medical Research and Development (AMED). 

\section*{References}
\bibliographystyle{unsrt}
\bibliography{references}

\end{document}